\documentclass{iopart}

\expandafter\let\csname equation*\endcsname\relax
\expandafter\let\csname endequation*\endcsname\relax

\usepackage{lipsum} 
\usepackage{color}
\usepackage{graphicx}
\usepackage{float}
\usepackage{iopams}
\usepackage[margin=10pt,font=small,format=hang]{caption}
\usepackage{amsmath,amssymb}
\usepackage{acronym}
\usepackage{multirow}
\usepackage{tabu}
\usepackage{tikz}
\usetikzlibrary{arrows,shapes,trees,decorations.pathreplacing}
\usepackage{import}
\usepackage{svn-multi}
\usepackage{lineno}
\usepackage{hyperref}
\usepackage{mathtools}
\usepackage{xspace}
\usepackage{tabularx}
\usepackage{url}
\usepackage{cite}
\hypersetup{
    colorlinks=true,
    linkcolor=blue,
    filecolor=magenta,      
    urlcolor=cyan,
}

\DeclareGraphicsExtensions{%
    .pdf,.PDF,%
    .png,.PNG,%
    .jpg,.mps,.jpeg,.jbig2,.jb2,.JPG,.JPEG,.JBIG2,.JB2}

\newcommand{\result}[1]{#1}

\newcommand{\bw}{\texttt{BayesWave}\xspace}
\newcommand{\gs}{\texttt{gwsubtract}\xspace}

\def\mystrut{\vrule height 9.3pt depth 3.1pt width 0pt}

\makeatletter
\newcommand\footnoteref[1]{\protected@xdef\@thefnmark{\ref{#1}}\@footnotemark}
\makeatother

\begin{document}

\title[]{Subtracting glitches from gravitational-wave detector data during the third observing run}
\author{D Davis$^1$,
        T B Littenberg$^2$,
        I M Romero-Shaw$^{3,4,5}$,
        M Millhouse$^6$,
        J McIver$^7$,
        F Di Renzo$^{8,9}$,
        G Ashton$^{10}$
        }

\address{$^1$LIGO, California Institute of Technology, Pasadena, CA 91125, USA}
\address{$^2$NASA Marshall Space Flight Center, Huntsville, AL 35811, USA }
\address{$^3$OzGrav, School of Physics \& Astronomy, Monash University, Clayton 3800, Victoria, Australia}
\address{$^4$Monash Astrophysics, School of Physics and Astronomy, Monash University, VIC 3800, Australia}
\address{$^5$Department of Applied Mathematics and Theoretical Physics, Cambridge CB3 0WA, United Kingdom}
\address{$^6$OzGrav, University of Melbourne, Parkville, Victoria 3010, Australia}
\address{$^7$University of British Columbia, Vancouver, BC V6T 1Z4, Canada}
\address{$^8$Universit\`a di Pisa, I-56127 Pisa, Italy}
\address{$^9$INFN, Sezione di Pisa, I-56127 Pisa, Italy}
\address{$^{10}$Department of Physics, Royal Holloway, University of London, TW20 0EX, United Kingdom}
\date{\today}

\begin{abstract}
Data from ground-based gravitational-wave detectors contains numerous short-duration instrumental artifacts, called ``glitches.''
The high rate of these artifacts in turn results in a significant fraction of gravitational-wave signals from compact binary coalescences overlapping glitches. 
In LIGO-Virgo's third observing run, $\approx 20\%$ of signals required some form of mitigation due to glitches. 
This was the first observing run that glitch subtraction was included as a part of LIGO-Virgo-KAGRA data analysis methods for a large fraction of detected gravitational-wave events. 
This work describes the methods to identify glitches, the decision process for deciding if mitigation was necessary, and the two algorithms, \bw and \gs, that were used to model and subtract glitches. 
Through case studies of two events, GW190424\_180648 and GW200129\_065458, we evaluate the effectiveness of the glitch subtraction, compare the statistical uncertainties in the relevant glitch models, and identify potential limitations in these glitch subtraction methods. 
We finally outline the lessons learned from this first-of-its-kind effort for future observing runs. 
\end{abstract}

\maketitle

\section{Introduction}\label{sec:intro}

Data analysis of gravitational-wave signals from compact binary coalescences (CBCs)~\cite{GWTC-1,GWTC-2,GWTC-3, GWTC-2.1} requires precise understanding of the expected astrophysical waveforms as well as models of the noise in gravitational-wave detectors~\cite{LIGOScientific:2019hgc}. 
These analyses, and the identification of times containing gravitational waves, are limited by the amplitude of background noise, referred to as the sensitivity of the detectors. 
Numerous noise sources limit the sensitivity of ground-based gravitational-wave detectors~\cite{aligo, avirgo, kagra}, including fundamental noise sources that are due to the physical limitations of the design of the detectors, and technical noise sources, that are due to the imperfections in the current detectors~\cite{aLIGO:2020wna, Acernese:2022jes, KAGRA:2022fgc}. 
There are also short bursts of excess noise in the detectors, referred to as ``glitches,'' that further pollute the gravitational-wave strain data~\cite{LIGO:2021ppb,Acernese:2022jes,KAGRA:2020agh}. 

Typical Bayesian methods in gravitational-wave data analysis assume that the noise is stationary and Gaussian over the time period analyzed~\cite{LIGOScientific:2019hgc}.
These assumptions are violated, however, when glitches are present in the analyzed data. 
When not accounted for as a part of the analyses, the presence of glitches can bias estimates of gravitational-wave properties~\cite{Powell:2018csz,Kwok:2021zny,Mozzon:2021wam,Macas:2022afm}.
There are some techniques proposed in which these assumptions have been relaxed~\cite{Talbot:2021igi, Chatziioannou:2021ezd, Capano:2021etf,Kumar:2022tto}, but these methods have not yet been used in an analysis by the LIGO-Virgo-KAGRA (LVK) collaboration.

Rather than being a rare edge case, it is quite common for glitches to be present near a gravitational-wave signal. 
The median rate of glitches in the LIGO and Virgo detectors during their third observing run (O3)~\cite{GWTC-2,GWTC-3} were above \result{1 per minute} for the majority of the run. 
At this rate, the probability a glitch to overlap a signal from a binary black hole is $\mathcal{O}(10\%)$ and it is almost certain that a glitch will overlap a signal from a binary neutron star.
These pessimistic estimates have held true for recent catalogs.
In total, 18 of the events detected in O3 were impacted by nearby glitches~\cite{GWTC-2,GWTC-3,GWTC-2.1}, including all confidently detected gravitational-wave signals from binary neutron stars~\cite{GW170817,GW190425} and neutron star -- black hole binaries~\cite{GW190814, NSBH}. 
Since typical methods used to estimate gravitational-wave source properties do not account for these glitches, it is therefore important to develop alternate mitigation methods to reduce any potential biases. 

Three main categories of glitch mitigation have been used in analyses of CBCs: removing the entire stretch of data containing a glitch; modelling glitches using the strain data and subtracting this model; and using an additional time series that witnesses the source of the glitch to subtract the relevant excess noise. 
The first approach is commonly used in searches for CBCs~\cite{Usman:2015kfa,Zackay:2019kkv}, but requires additional modifications to the analysis techniques to be used in parameter estimation without introducing biases~\cite{Pankow:2018qpo,Capano:2021etf}.
For this reason, only the latter two techniques have been used in analyses by the LVK.

A wide variety of similar methods have been developed to model and subtract glitches. 
Many techniques have been developed to model glitches using gravitational-wave strain data alone, including modeled Bayesian approaches~\cite{Cornish:2014kda,Pankow:2018qpo,Cornish:2020dwh,Chatziioannou:2021ezd, Hourihane:2022doe,Merritt:2021xwh} and some based on machine learning~\cite{Wei:2019zlc,Torres-Forne:2020eax}. 
Likewise, techniques that utilize the large number of additional auxiliary sensors at each observatory to model sources of noise with analytic methods\cite{Allen:1999ct,Driggers:2012noise,Meadors:2014,Tiwari:2015,Driggers:2017,Davis:2019,Was:2020ziy,T2100058,VIRGO:2021kfv,KAGRA:2022frk} and machine-learning approaches~\cite{Vajenta:2020ml,Ormiston:2020ele,Mukund:2020lby,Yu:2021swq,Mogushi:2021deu} have been proposed.

Of these tools, the only algorithms that have been used to model and subtract glitches for LVK analyses~\cite{GWTC-1,GWTC-2,GWTC-3,GWTC-2.1} are \bw~\cite{Cornish:2014kda,Pankow:2018qpo,Cornish:2020dwh} and \gs~\cite{Allen:1999ct,Davis:2019}.
\bw models glitches based on only the gravitational-wave strain data using wavelets.
It has been used to subtract glitches in LVK analyses since the second observing run (O2)~\cite{GW170817, Pankow:2018qpo} and is the most commonly used algorithm to date.
Conversely, \gs was used to subtract glitches in an LVK analysis for the first time in O3~\cite{GWTC-3}. 
However, this method was previously used for broadband noise subtraction in O2~\cite{Davis:2019}. 

This work focuses on describing the versions of these tools that were used in O3, the methods for determining if glitch subtraction was needed, and the potential limitations of current glitch subtraction methods.
We first provide background on these two methods in Section~\ref{sec:methods}, highlighting how the specific configuration of the tools used in O3 differs from previous or more recent versions of the tools. 
Section~\ref{sec:methods} also explains the methods used to evaluate the need for and results of glitch subtraction.
We then focus on two events, GW190424\_180648 and GW200129\_065458 (which we will refer to as GW190424 and GW200129), in Section~\ref{sec:real_data} and investigate the result of glitch subtraction in each case and estimate the additional statistical errors introduced from glitch modelling. 
Finally, in Section~\ref{sec:conc}, we describe potential updates to these glitch mitigation methods for the next observing run and the expected challenges for glitch subtraction with a higher detection rate. 

\section{Methods}\label{sec:methods}

In the third observing run, two glitch subtraction methods were utilized as a pre-processing step before parameter estimation analyses were completed~\cite{GWTC-2,GWTC-3,GWTC-2.1}. 
The first algorithm, \bw~\cite{Cornish:2014kda,Pankow:2018qpo,Cornish:2020dwh}, was used for \result{15} events in O3~\cite{GWTC-2,GWTC-3,GWTC-2.1}. 
The second algorithm, \gs~\cite{Davis:2019}, was used for only \result{1} event~\cite{GWTC-3}. 
This section outlines the basic methods both algorithms use to model the nose contributions from a glitch. 
We also discuss the metrics we used to evaluate if glitch subtraction was needed and if the glitch subtraction was successful at removing the excess power from any identified glitches. 

\subsection{Glitch subtraction using wavelets}\label{sec:bw_sub}

The \bw algorithm assumes that in each interferometer, the detector data $h(t)$ can be expressed as 
\begin{equation}
  h(t) = n(t) + s(t) + g(t)
  \label{Eq:BWNoiseModel}
\end{equation}
where $n(t)$ is Gaussian noise, $s(t)$ is an astrophysical signal, and $g(t)$ is a non-Gaussian instrumental glitch.
\bw models non-Gaussian features (both signal and glitch) as a sum of sine-Gaussian (also called Morlet-Gabor) wavelets.
In the time domain, the wavelets are given by
\begin{equation}
  \Psi(t;\vec{\theta}) = A e^{-(t-t_0)^2/\tau^2}\cos\left(2\pi f_0(t-t_0)+\phi_0\right)
  \label{Eq:Wavelet}
\end{equation}
where $\vec{\theta}=\{t_0, f_0, Q, A, \phi_0\}$ are the parameters of each wavelet, and $\tau=Q/(2\pi f_0)$.
The wavelet parameters, as well as the total number of wavelets used, is marginalised over via a transdimensional Markov chain Monte Carlo.

For the glitch model, the wavelet parameters in each detector are independent.
For the signal model, there is one common set of wavelets which are projected onto the detectors using a set of extrinsic parameters that describe the sky location and polarisation content.
The power spectral density is modeled using~\cite{Littenberg:2014oda}, which also uses a transdimensional MCMC to model the noise as a combination of a cubic spline and Lorentzians.
Full details can be found in~\cite{Cornish:2014kda} and ~\cite{Cornish:2020dwh}.

The transient features in the data can be reconstructed using three different model assumptions:
\begin{itemize}
  \item[I.] The data contain Gaussian noise with an astrophysical signal.
  \item[II.] The data contain Gaussian noise with an instrumental glitch.
  \item[III.] The data contain Gaussian noise with both a signal \emph{and} a glitch.
\end{itemize}

To identify and remove glitches from detector data, we use either the glitch-only model (case II above), or signal-plus-glitch model (case III).
The glitch-only approach is used when the glitch is sufficiently separated in time and/or frequency from the part of the signal that \bw can reconstruct.
When using the glitch-only approach, we only need to consider data from the single detector in which the glitch appears.
If the signal and glitch overlap significantly, we use the signal-plus-glitch model. 
The signal-plus-glitch model requires data from multiple detectors in order to separate the coherent signal power and the incoherent glitch power.
By simultaneously fitting the for signals and glitches, we ensure that we do not subtract any signal power during glitch mitigation.

As \bw uses an MCMC to marginalise over the parameters, the end result is a posterior distribution of time series of the glitch, $g(t)$.
To actually perform the glitch mitigation, we select a fair draw from the posterior~\cite{Hourihane:2022doe}, and subtract it from the detector data.
The code for both \bw and the \bw glitch subtraction methods can be found at~\cite{BWrepository}.

\subsection{Linear subtraction with a witness}\label{sec:lin_sub}

The codebase used to subtract glitches using an auxiliary witness, \gs, is the same linear subtraction algorithm as was presented in~\cite{Davis:2019}. 
In this section, we provide a short description of this method, 
and note how the use case considered in this work differs from the original use case presented in~\cite{Davis:2019}.
The code associated with \gs is publicly available at~\cite{gwsubtract}.

In order to subtract noise, we first assume that the measured strain in the gravitational-wave detector, $h(t)$, is a linear combination of time series from different sources, $n_i(t)$. 
We further assume that at least one of these noise sources can be modelled as the convolution of a witness time series, $a(t)$, and an unknown transfer function $c_{ah}(t)$.
In the time domain, the gravitational-wave strain can therefore be written as 
\begin{linenomath}\begin{equation}
\begin{split}
h(t) &= n_1(t) + n_2(t) + \ldots + n_N(t) \\
&= n_1(t) + n_2(t) + \ldots + a(t) * c_{ah}(t) \\
&= h'(t) + a(t) * c_{ah}(t)
\end{split}
\end{equation}\end{linenomath}
where all noise sources except one are included in $h'(t)$.
This can be conveniently cast into the frequency domain as
\begin{linenomath}\begin{equation}
\Tilde{h}(f) = \Tilde{h}'(f) + \Tilde{a}(f) \cdot \Tilde{c}_{ah}(f)
\end{equation}\end{linenomath}

Following the derivation from~\cite{Allen:1999ct, Davis:2019}, we can estimate the unknown transfer function by
using the discrete Fourier transform of the strain data, $\Tilde{Y}_h(f)$, and the witness time series, $\Tilde{Y}_h(f)$. 
A single value is calculated for the frequency band $[f_1, f_2)$ via
\begin{linenomath}\begin{equation}
\Tilde{c}_{ah}(f') = \frac{df}{f_2-f_1} \sum_{f=f_1}^{f_2} \Tilde{Y}_h(f) \Tilde{Y}_a^*(f),
\end{equation}\end{linenomath}
where $df$ is the frequency resolution of the data.  

By averaging nearby frequencies when calculating the transfer function, contributions to the measurement from chance correlations are reduced. 
Specifically, the fractional measurement error from this averaging method is expected to be
$\sqrt{\frac{df}{f_2-f_1}}$ if both $h(t)$ and $a(t)$ are stationary Gaussian noise. 

Although the basic description of the process for glitch subtraction is quite similar to~\cite{Allen:1999ct, Davis:2019}, 
the process has a few practical differences. 
In~\cite{Allen:1999ct, Davis:2019}, multiple witness time series were used to subtract contributions from a single noise source, requiring that correlations between each of the witness time series were also calculated. 
In this work, only a single witness time series is considered. 

An additional key difference is the stationary nature of the time series. 
Data that contains glitches is, by definition, not a stationary Gaussian time series. 
Regardless of the use case, if the transfer function is still stationary and linear, then theoretically this should not prevent these linear subtraction methods from being used. 
However, it is possible that the true transfer functions may have some non-linear features that are not captured by this method.
It is also possible that some of the methods used in~\cite{Davis:2019} to reduce the impact of false correlations could prevent the transfer function from being measured accurately. 
For these reasons, the ideal type of glitch to subtract with these methods is one that occurs at a high rate over a short time period. 
This increases the probability that the transfer function will be accurately measured. 

Due to these practical differences between glitch subtraction and the use case that~\cite{Davis:2019} was originally designed for, we completed additional validation tests to confirm that the methods of~\cite{Davis:2019} were able to be used for glitch subtractions. 
These tests are summarized in~\ref{sec:sim_data}.

\subsection{Estimating the significance of residuals}\label{sec:sig}

\begin{figure}[t]
  \centering
  \includegraphics[width=\textwidth]{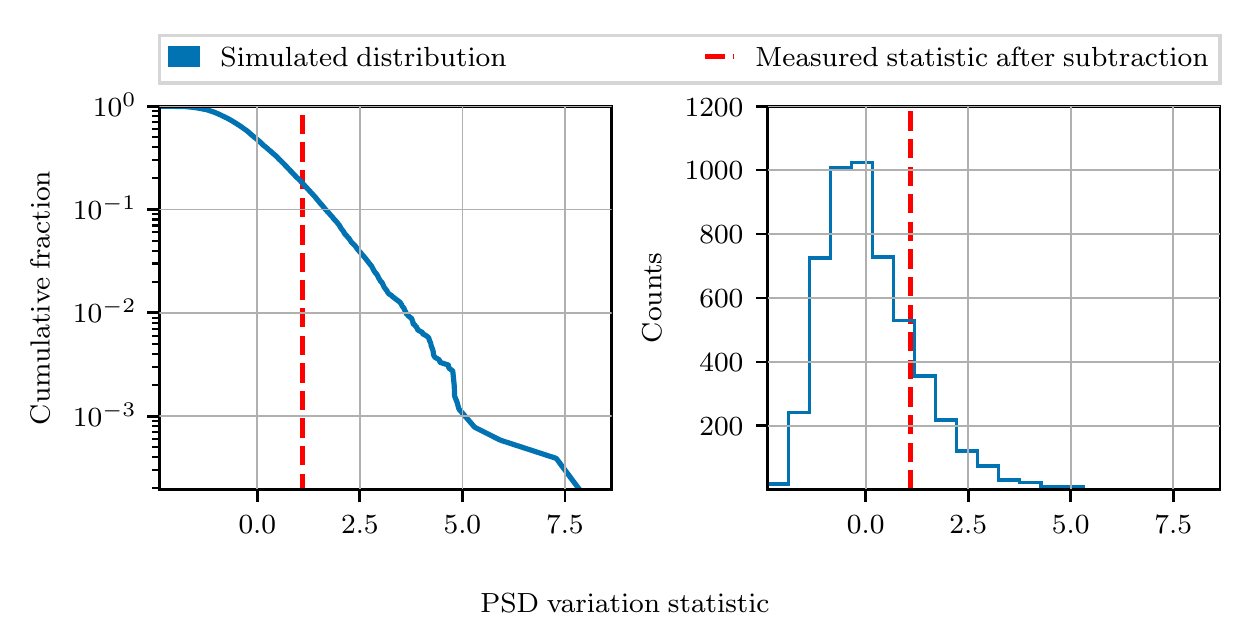}
  \caption{An example of the distribution of the stationarity statistic that is used to calculate the p-value of any excess power identified in the data. This example uses data from around GW190425~\cite{GW190425} from LIGO Livingston. The distribution of the statistic over the simulated background is shown in blue and the measured statistic after glitch mitigation is marked in red. Before subtraction, the statistic is much larger than any trial in the simulated background (and is not shown on this plot). 
  After subtraction, the measured statistic is consistent with the background.}
  \label{fig:pval}
\end{figure}

One of the key steps in the glitch subtraction process, after deciding whether to subtract excess power in the data, is determining whether this excess power was successfully subtracted. 
For these purposes, we employ a residual test inspired by~\cite{Mozzon:2020gwa} to evaluate the statistical significance of any non-Gaussian features in the data. 
This test relies upon comparing the power spectral density (PSD) of the data during the time of interest against the PSD measured over \result{512\,seconds}. 
Any positive fluctuations in the short-duration PSD are considered excess power.

Following the description in~\cite{Mozzon:2020gwa}, we first estimate the overall level of fluctuation in the PSD using \texttt{GWpy}~\cite{2021SoftX..1300657M}. 
We measure the PSD, $S_G(f)$, at the time of interest with a window the same size as the duration of the potential glitch. 
We then measure the PSD over \result{512 seconds} of data, $S_L(f)$, using the ``median method''~\cite{Allen:2005fk} with windows the same size as the short duration PSD and \result{$50\%$ overlap}.
The variation in the PSD, $\rho$ is given by 
\begin{linenomath}\begin{equation}
    \rho = \sum_f S_L(f) \times \sum_f \frac{S_G(f)}{S_L^2(f)} .
\end{equation}\end{linenomath}

However, the variance of this statistic is dependent on both the time window used to estimate the PSD, $\Delta t$, and the ``effective bandwidth'' considered, $\Delta f_{\rm eff}$~\cite{Mozzon:2020gwa}, making it unsuitable for comparisons where these two quantities differ. 
Specifically, the variance is 
\begin{linenomath}\begin{equation}
    \mathrm{Var}(\rho) = (\Delta t \Delta f_{\rm eff})^{-1} .
\end{equation}\end{linenomath}
The effective bandwidth is dependent on both the bandwidth considered, $\Delta f = f_{\rm high}-f_{\rm low}$, and the shape of the PSD. 
This effective bandwidth is calculated via 
\begin{linenomath}\begin{equation}
    \Delta f_{\rm eff} \approx \int_{f_{\rm low}}^{f_{\rm high}} \frac{{\rm max} [S_L(f)]}{S_L(f)}df .
\end{equation}\end{linenomath}
To account for this dependence on $\Delta t$ and $\Delta f$, we normalize the statistic by the standard deviation for the parameters we use in each instance and subtract the mean of the distribution. 
Hence our new statistic, $\Tilde{\rho}$ is 
\begin{linenomath}\begin{equation}
    \Tilde{\rho} = (\rho - 1) \times \sqrt{\Delta t \Delta f_{\rm eff}} .
\end{equation}\end{linenomath}
This new statistic has a mean of $0$ and a variance of $1$ for all choices of $\Delta t$ and $\Delta f$.

Although this new statistic is less susceptible to analysis choices for the reasons listed above, the true variance of the statistic may be different due to practical limitations in the measurement of the PSD. 
For example, it is not possible to calculate this statistic when $\Delta f < \Delta t^{-1}$ as there will be no points in the measured PSD in the bandwidth of interest. 
When $\Delta f \approx \Delta t^{-1}$, we also see additional variance in the statistic that is not accounted for in the normalization due to the large frequency resolution of the PSD.
If the frequency resolution of the PSD is sufficiently coarse, only a small number of data points are the bandwidth of interest, inflating the measured variance. 

To address this issue, we calculate a p-value for the measured value of $\Tilde{\rho}$ for a specific choice of $\Delta t$ and $\Delta f$ using simulated data. 
In each case, we generate \result{2048 seconds} of stationary, Gaussian data with the measured long-duration PSD using \texttt{Bilby}~\cite{Ashton:2018jfp, Romero-Shaw:bilby:2020}.
We then repeat the same procedure over each sequential short-duration time window for the whole time period simulated.
These results are used to generate the distribution from which the p-value is estimated. 

An example distribution of this statistic based on data around GW190425~\cite{GW190425} is shown in Figure~\ref{fig:pval}. 
Before subtraction, the measured statistic value was $\approx180$, much larger than any trial in the simulated background. 
After subtraction, the statistic dropped to $1.4$, which was higher than $18\%$ of the simulated background, corresponding to a p-value of 0.18. 
This can be seen from the cumulative distribution of the simulated background, shown on the left, and a histogram of the measured statistics in the background, shown on the right. 
The plotted histogram also demonstrates that the peak of this distribution is indeed close to 0. 
The distribution of statistics is asymmetric, with a longer tail towards larger values.
This asymmetry, even for cases that should be well-behaved, motivates the use of the p-value rather than a threshold on the measured statistic based on an assumed variance.

\subsection{Glitch subtraction validation procedures}\label{sec:process}

In order to ensure that the glitch-subtracted data used in downstream analyses did not introduce biases, a series of glitch-subtraction review tests were completed as a part of the O3 LVK analyses~\cite{GWTC-2,GWTC-3, GWTC-2.1}. 
These tests focused on ensuring that either the data was consistent with Gaussian noise or that any residual instrumental excess power would not bias estimates of an event's source properties.
Previous investigations into the potential biases on parameter estimation when glitches overlap or are nearby signals~\cite{Powell:2018csz,Kwok:2021zny,Mozzon:2021wam,Macas:2022afm} have not yet covered the entire parameter space of glitch types and where in the signal the glitch overlaps. We therefore used conservative thresholds for each test to minimize potential biases.

In order to identify if data surrounding gravitational-wave events required glitch subtraction, we first utilized the results of event validation procedures completed for each event.
These procedures for O3 are detailed in~\cite{Davis:2020nyf, Acernese:2022jes}.
The outcome of event validation specifically indicated if any glitches were nearby a gravitational-wave signal. 
Events that did not have any glitches flagged by these procedures were deemed not to require glitch subtraction. 

For flagged events, the residual test described in Section~\ref{sec:sig} was used to determine if the data was inconsistent with Gaussian noise. 
The time-frequency bounds to use for the residual test were determined by visual inspection of the data around events using spectrograms produced with the Q-transform~\cite{Chatterji:2004qg}.
Cases where the residual p-value was greater than \result{$0.1$} were deemed consistent with Gaussian noise and no glitch subtraction was attempted, 
cases with p-values less than \result{$0.001$} were passed on for glitch subtraction, 
and cases with p-values between these two thresholds required further follow up before a decision was reached. 
In the cases when the result was inconclusive, the overlap of the excess power and signal was used to determine if glitch subtraction was needed; glitch subtraction was generally recommended in cases where excess power directed overlapped the signal. 
As previously described, glitch subtraction was completed with \bw in all but \result{1} case where \gs was used. 

Once the glitch subtraction was completed, the data was reevaluated to determine if the excess power was sufficiently subtracted. 
The residual metric was recomputed for the glitch-subtracted data using the same time-frequency regions.
The same p-value thresholds were also used to decide if the data was now consistent with Gaussian noise. 

In addition to the residual tests, comparisons of parameter estimation results using different mitigation methods were used to evaluate if further glitch subtraction was needed or if a different glitch mitigation method should be used\footnote{These parameter estimation comparison tests were completed when practical for all events that failed the initial residual test, regardless if the glitch-subtracted data passed the residual test.}.
To test if excess power in the data could bias analyses, we estimated the source properties of events using either the nominal lower frequency limit (20\,Hz) or a higher lower limit that excluded the data containing glitches. 
In cases where the glitch subtraction was able to improve the stationarity of the data, analyses using the original and glitch subtracted data were also compared. 
If the glitch-subtracted data did not pass the residual test, different frequency bounds that excluded the flagged time-frequency regions were recommended if the different frequency bound changed the measured chirp mass\footnote{\label{foot:params}See~\cite{Romero-Shaw:bilby:2020} and references therein for definitions of these parameters.} posterior by more than $1\sigma$ and that the signal-to-noise ratio (SNR) was reduced by less than $10\%$. 
Other parameters, such as $\chi_p$ and $\chi_\text{eff}$\footnoteref{foot:params} were also compared between analyses.
Although not the deciding factor in determining which mitigation method was recommended, any differences in these parameters between analyses were flagged. 
Additional details about these tests can be found in~\ref{sec:pe}.

Once either the data around an event passed the residual test or the identified glitches were deemed to not bias parameter estimation, 
the data was approved for use in downstream analyses. 
These procedures generally were completed for each event within a few days (for events not requiring glitch subtraction) to weeks (for events that required glitch subtraction). 
The timeline for this process was limited by the need for human input at multiple steps; in future observing runs, additional automation will be needed to significantly speed up these procedures.

\section{Results with real events}\label{sec:real_data}

In this section, we will consider the data around two events, GW190424~\cite{GWTC-2} and GW200129~\cite{GWTC-3}.
Both events directly overlapped glitches that were mitigated using the glitch subtraction procedures discussed in this work. 
Data around GW190424 was processed with \bw and data around GW200129 was processed using \gs.
GW190424 was one of the more challenging cases of glitches that were subtracted with \bw while GW200129 is the only event in O3 processed with \gs.  
In both cases, however, there are well-known witnesses to the source of the glitches near each event. 
We are therefore confident that the power subtracted is due to glitches as opposed to an unmodelled astrophysical signal. 

GW190424 directly overlapped two different types of glitches; one glitch was due to a mechanical camera shutter that was inadvertently left running on a timer~\cite{Davis:2020nyf} and one was due to scattered light that is reflected off a moving surface~\cite{Soni:2020rbu}.
A data quality flag was developed that removed time segments containing glitches from the astrophysical searches due to these camera shutter glitches~\cite{T2100045}.
GW190424 is an event of interest as one of the signals in O3 that was detected using only the data from a single gravitational-wave detector. 
There was therefore no data from a different detector available to help disentangle coherent power (that is astrophysical) from incoherent power (that is instrumental). 
Instead, glitch power was identified using auxiliary witnesses and then modelled using \bw.
For events in O3 that were detected in more than one detector, it was instead possible to disentangle incoherent and coherent power using the multiple strain time series. 

GW200129 directly overlapped glitches that were due to a fault in the 45 MHz electro-optic modulator driver system at LIGO Livingston~\cite{GW150914_detchar}.
This type of glitch was repeating for many minutes around the event and was well-witnessed by the modulation control system. 
For these reasons, this glitch type was able to be addressed using linear subtraction.
These glitches were also vetoed by a data quality flag in O3~\cite{T2100045}. 
GW200129 is of interest astrophysically as the gravitational-wave event with potentially the strongest evidence to date in favor of spin-induced preccession in a CBC~\cite{GWTC-3,Hannam:2021pit,Varma:2022pld,Payne:2022}. 
The gravitational-wave strain from a spin-precessing binary is lower at specific frequencies as compared to the corresponding non-precessing signal.
As the effect of orbital precession could be mimicked by the presence of non-Gaussian noise, it is essential to the analysis of any signal that any glitches present are well-modelled and subtracted. 
Conversely, the uncertainties on glitch modelling could introduce additional uncertainties in the parameter estimation results~\cite{Payne:2022}. 

Including these two events, a total of \result{16} events required glitch subtraction in O3. Table~\ref{tab:glitch_sub} lists all of these events, the interferometer from which the glitch was subtracted, and the time-frequency regions that were targeted for glitch subtraction. 
Note that due to the nature of the glitch subtraction algorithms used, glitch power may also have been subtracted from outside the listed time-frequency regions. 
However, these time-frequency bounds correspond to the time windows and bandwidth targeted for glitch subtraction and used for residual tests. 
In addition to these events, further \result{10} events\footnote{GW190720\_000836, GW190828\_065509, GW190910\_112807, GW190915\_235702, GW191204\_171526, GW200208\_130117, GW200219\_094415, GW200220\_124850, GW200311\_115853, and GW200220\_124850} were assessed with these review processes but it were determined to not required either due to the identified time-frequency regions passing the residual test or being located outside the time window used in parameter estimation analyses.  
All glitch-subtracted data is publicly available~\cite{GWTC2_subdata,GWTC3_subdata,GWTC2.1_subdata}.

\begin{table}
\begin{tabularx}{\linewidth}{l l l r r}
\textbf{Event} & \textbf{GPS time} & \textbf{IFO} & \textbf{Time (s)} & \textbf{Frequency (Hz)} \\
\hline \hline
GW190413\_134308 & 1239198206.74 & L & [-1.7, 0.7] & [20, 35] \\
\makebox[0pt][l]{\fboxsep0pt\colorbox{lightgray}{\mystrut\hspace*{0.955\linewidth}}}\!\! GW190424\_180648 & 1240164426.14 & L & [-.3, 0.1] & [60, 100] \\
\makebox[0pt][l]{\fboxsep0pt\colorbox{lightgray}{\mystrut\hspace*{0.955\linewidth}}}\!\!  &  &  & [-1.5, 0.5] & [20, 40] \\
GW190425 & 1240215503.02 & L & [-61.57, -59.97] & [20, 50] \\
\makebox[0pt][l]{\fboxsep0pt\colorbox{lightgray}{\mystrut\hspace*{0.955\linewidth}}}\!\! GW190503\_185404 & 1240944862.30 & L & [-1.1, 0.5] & [20, 35] \\
\makebox[0pt][l]{\fboxsep0pt\colorbox{lightgray}{\mystrut\hspace*{0.955\linewidth}}}\!\!  &  & L & [-3.9, -1.9] & [20, 40] \\
GW190513\_205428 & 1241816086.75 & L & [-5.65, 0.15] & [15, 40] \\
\makebox[0pt][l]{\fboxsep0pt\colorbox{lightgray}{\mystrut\hspace*{0.955\linewidth}}}\!\! GW190514\_065416 & 1241852074.89 & L & [-1.3, 2.7] & [15, 30] \\
\makebox[0pt][l]{\fboxsep0pt\colorbox{lightgray}{\mystrut\hspace*{0.955\linewidth}}}\!\!  &  & L & [-5.3, 0.3] & [15, 30] \\
GW190701\_203306 & 1246048404.80 & L & [-1.28, 1.72] & [20,55] \\
\makebox[0pt][l]{\fboxsep0pt\colorbox{lightgray}{\mystrut\hspace*{0.955\linewidth}}}\!\! GW190924\_021846 & 1253326744.85 & L & [-5.0, -3.0] & [18,110] \\
GW191105\_143521 & 1256999739.93 & V& [-7.05, -5.65] & [25, 50] \\
 &  &  & [0.1, 7.1] & [15, 30] \\
 &  &  & [-4.35, -3.15] & [16, 22] \\
 &  &  & [-6.2, -5.0] & [18, 25] \\
\makebox[0pt][l]{\fboxsep0pt\colorbox{lightgray}{\mystrut\hspace*{0.955\linewidth}}}\!\! GW191109\_010717 & 1257296855.22 & H & [-3.1, -0.1] & [25, 45] \\
\makebox[0pt][l]{\fboxsep0pt\colorbox{lightgray}{\mystrut\hspace*{0.955\linewidth}}}\!\!  &  & L & [-1.5, 1.5] & [20, 32] \\
GW191113\_071753 & 1257664691.84 & H & [-1.75, 1.75] & [25, 60] \\
 &  &  & [-4.4, -3.6] & [25, 55] \\
\makebox[0pt][l]{\fboxsep0pt\colorbox{lightgray}{\mystrut\hspace*{0.955\linewidth}}}\!\! GW191127\_050227 & 1258866165.55 & H & [-2.125, 0.375] & [25, 40 \\
\makebox[0pt][l]{\fboxsep0pt\colorbox{lightgray}{\mystrut\hspace*{0.955\linewidth}}}\!\!  &  &  & [0.0, 3.0] & [20, 45] \\
GW191219\_163120 & 1260808298.45 & H & [-7.85, -4.35] & [20, 1000] \\
 &  & L & [4.5, 7.0]  & [20, 25] \\
\makebox[0pt][l]{\fboxsep0pt\colorbox{lightgray}{\mystrut\hspace*{0.955\linewidth}}}\!\! GW200105\_162426 & 1262276684.06 & L & [-4.25, 0.25] & [18, 22] \\
GW200115\_042309 & 1263097407.74 & L & [-70, 30]  & [20, 25] \\
\makebox[0pt][l]{\fboxsep0pt\colorbox{lightgray}{\mystrut\hspace*{0.955\linewidth}}}\!\! GW200129\_065458 & 1264316116.44 & L & [-0.525, 0.275] & [30, 60] \\
\makebox[0pt][l]{\fboxsep0pt\colorbox{lightgray}{\mystrut\hspace*{0.955\linewidth}}}\!\!  &  &  & [0.6, 1.6] & [30, 55] \\
\end{tabularx}
\caption{A list of all of the events from O3 LVK analyses~\cite{GWTC-2,GWTC-3, GWTC-2.1} which required glitch subtraction. For each event, we list: the GPS time of the event; the interferometer(s) which required glitch subtraction; and the time window (relative to the event time) and frequency range targeted for glitch subtraction. These time-frequency regions were used for the residual tests of glitch-subtracted data. Note that additional excess power may have been subtracted outside of these time-frequency bounds. Glitches around all events were subtracted with \bw except for glitches around GW200129\_065458, which were subtracted with \gs. }
\label{tab:glitch_sub}
\end{table}

\subsection{Glitch models for events in O3}\label{sec:results}

The process of glitch subtraction can be broken into two stages: glitch modelling and glitch subtraction. 
The subtraction process is done simply by finding the difference between the glitch model and the original data. 
However, as described in Section~\ref{sec:methods}, the methods used used by the tools described in this work to estimate the glitch model differ significantly. 
These differences are highlighted in the two representative events we have considered in this section. 

\begin{figure}[t]
  \centering
  \includegraphics[width=\textwidth]{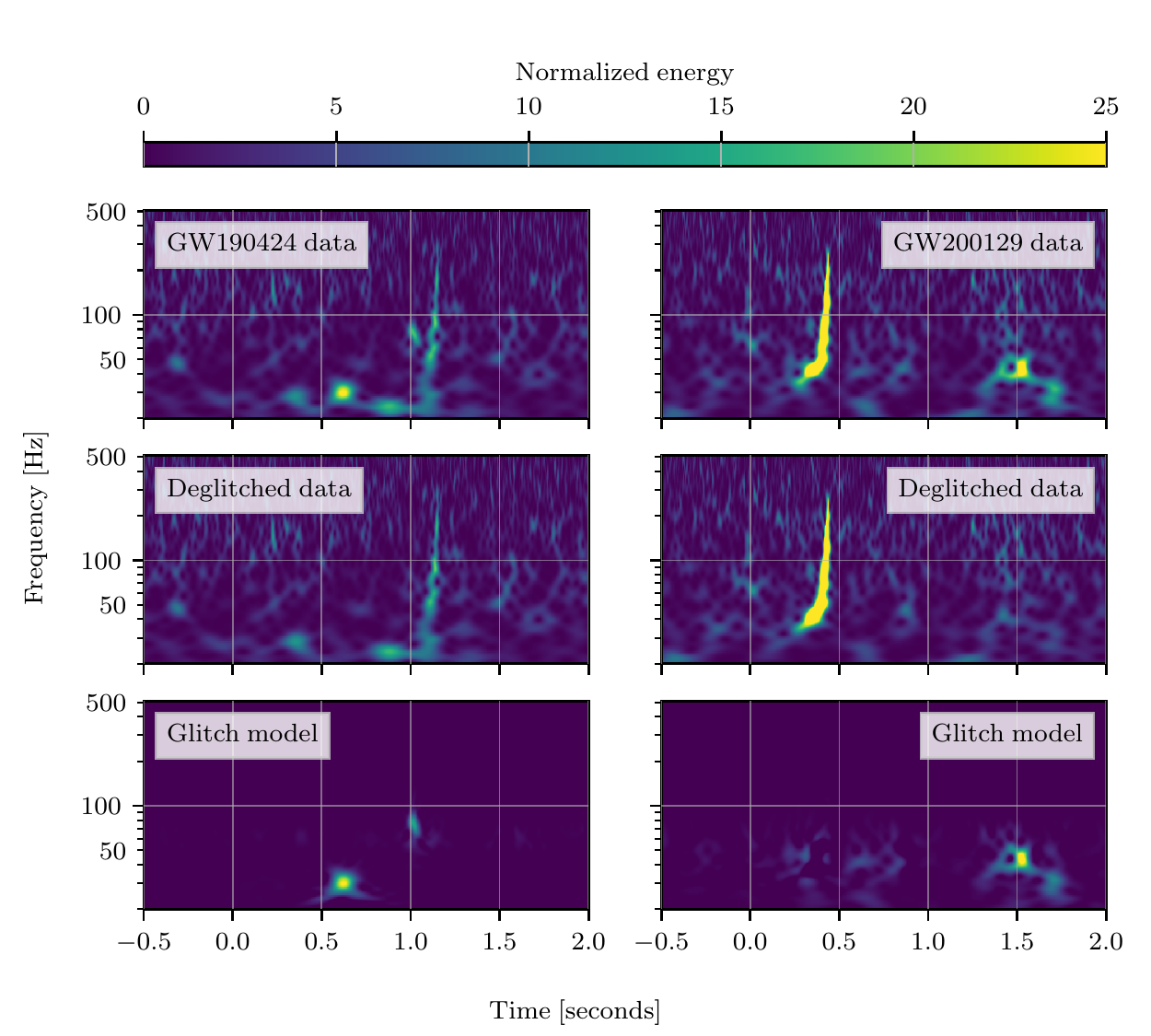} \caption{Spectrograms of the data around GW190424 and GW200129 before glitch subtraction (Top), the data after glitch subtraction (Middle), and the residual between the two  which corresponds to the subtracted glitch model (Bottom). Data around GW190424 is shown on the left and data around GW200129 is shown on the right. All spectrograms are generated with GWpy~\cite{2021SoftX..1300657M} using the Q-transform~\cite{Chatterji:2004qg}.
  The bottom panels highlight the time-frequency regions that were modified by glitch subtraction. 
  Glitches around GW190424 were processed using \bw, resulting to the subtraction distinct time-frequency regions. 
  Glitches around GW200129 were conversely processed with \gs, resulting in broadband subtraction below 70\,Hz. }
  \label{fig:omega}
\end{figure}

The first data visualizations we will consider for our comparisons are spectrograms generated using the Q-transform~\cite{Chatterji:2004qg}.
Spectrograms of the data around GW190424 and GW200129 from LIGO Livingston can be seen in Figure~\ref{fig:omega}. 
The panels of this figure show the data before subtraction, after subtraction, and the difference of the two spectrograms. 
The power subtracted near GW190424 is confined to two distinct burst of power: one corresponding to power from the camera shutter glitch at $\approx 70$\,Hz, and the other one from the arches in the nearby scattered light glitch. 
These bursts of power are not directly overlapping the signal, but are near enough that they might bias parameter estimation results. 
Conversely, the power subtracted near GW200129 is more broadband, with visible glitch power below 70\,Hz subtracted for multiple seconds that directly overlaps the signal.

\begin{figure}[t]
  \centering
  \includegraphics[width=\textwidth]{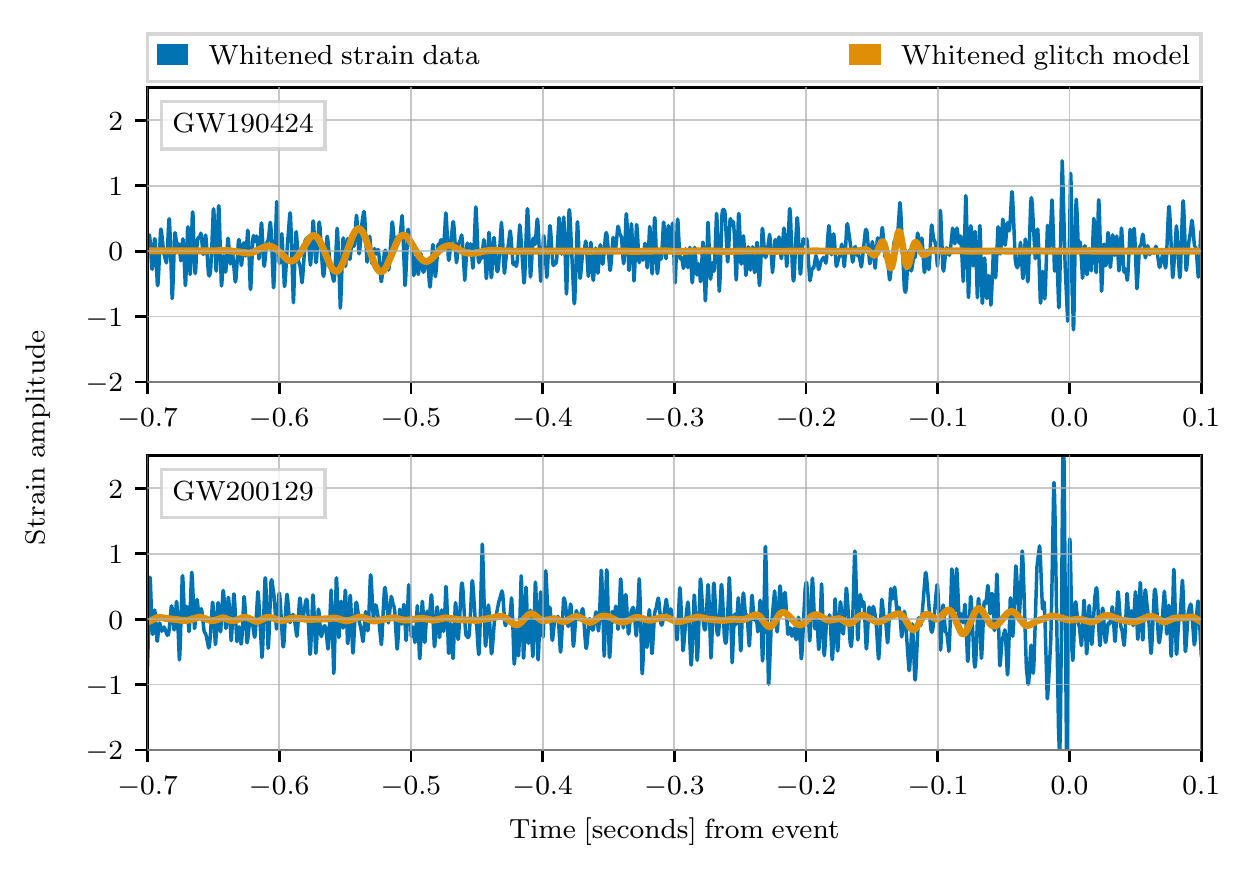}
  \caption{A plot of the whitened strain data from LIGO Livingston before glitch subtraction and the whitened glitch model that was subtracted near GW190424 and GW200129. The chirps corresponding to GW190424 and GW200129 are visible in the whitened strain data merging at $t=0$. In both the amplitude and frequency of the glitches near these events are comparable to amplitude and frequency of the signal. The glitch model for GW190424 is two distinct wavelets, while the glitch model for GW200129 is always non-zero over the plotted time segment.}
  \label{fig:glitch_wf}
\end{figure}

A comparison of the original strain data, the deglitched data, and the glitch models can be seen in Figure~\ref{fig:glitch_wf}.  
Both the glitch models and the strain data have been whitened using \texttt{GWpy}~\cite{2021SoftX..1300657M}. 
The same types of features that were visible in the Figure~\ref{fig:omega} can be seen in the whitened time series. 
The glitch model for GW190424 is confined to two short bursts, while the glitch model for GW200129 is non-zero over the entire time considered. 

These results also demonstrate limitations of these two methods. 
\bw was not able to successfully subtract all of the excess power from the fast scattering glitches, while the linear subtraction was not able to subtract any power above 70\,Hz. 
Scattered light glitches are a known challenge for \bw.
Recent investigations~\cite{Hourihane:2022doe} have shown that in order to fully model scattered light glitches with 
\bw, low-SNR wavelets must be used. 
These investigations also found that wavelets at the required SNRs are strongly disfavored when using priors similar to those used by the analyses discussed in this work. 
The linear subtraction method is also limited by the input data used to witness the glitch power. 
In the case of GW200129, the auxiliary sensor used in the subtraction was negligibly coherent with the gravitational-wave strain above 70\,Hz. 
In other cases, such as the broadband noise subtraction performed in O2~\cite{Davis:2019}, the sampling rate of the auxiliary sensors can limit the range of frequencies where subtraction is effective.
Ultimately the effectiveness of the linear subtraction method relies upon how faithfully the relevant auxiliary sensors are able to witness the noise source of interest.  

\subsection{Estimated uncertainties}\label{sec:errors}

As previously stated, the uncertainty in the model of a glitch overlapping a gravitational-wave signal introduces another potential source of uncertainty in the parameter estimation results. 
Although it is possible to sample over the glitch model as an additional component in the parameter estimation (as is possible with recent updates to \bw~\cite{Chatziioannou:2021ezd, Hourihane:2022doe}), this was not done in the O3 analyses. 
Instead, a single glitch model was subtracted from the data as a preprocessing step. 
Although this means that any glitch modelling errors are not included in the analysis, this is not expected to significantly bias the results if the chosen glitch model is at least representative of the true glitch model. 
In this subsection, we discuss how errors in glitch modelling have been assessed for both \bw and \gs, and demonstrate that statistical errors in the chose glitch models are qualitatively low.  

\begin{figure}[t]
  \centering
  \includegraphics[width=\textwidth]{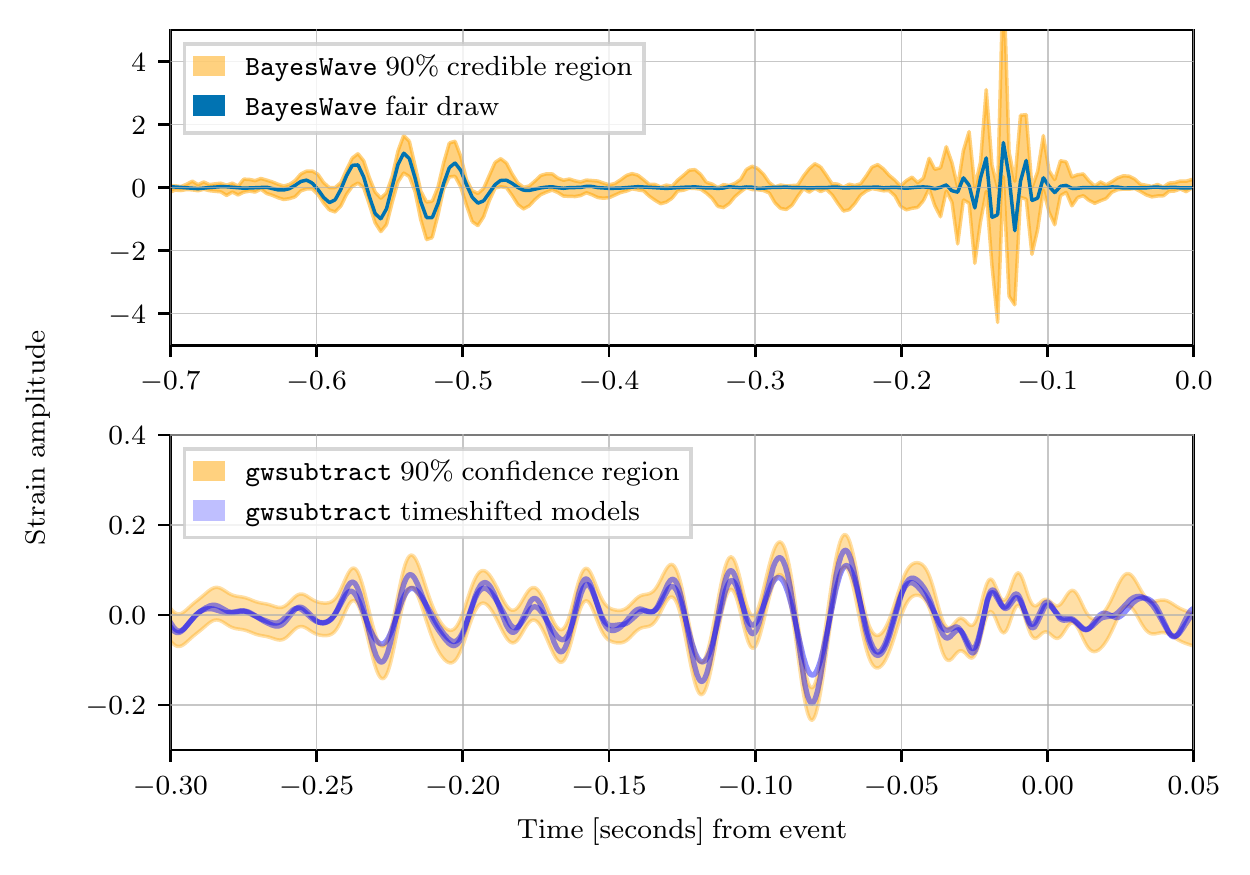}
  \caption{Top: The 90\% credible interval of the whitened glitch time series posterior from \bw, and a fair draw of a single whitened waveform from the posterior (blue).
  Bottom: A comparison of the 90\% confidence interval of the whitened glitch time series from \gs (orange) and the whitened glitch waveform subtracted when measuring the transfer function over different time periods (blue). The different time series are consistent in amplitude and phase.
  Note the differing axis scales for each panel. 
  The \bw 90\% credible is much larger than the plotted \gs 90\% confidence interval due to the different methods used by each algorithm to estimate the errors. Due to the differing error estimation methods, the relative amplitude of the shown credible/confidence regions should not be used to compare the precision of the two methods.}
  \label{fig:glitch_errors}
\end{figure}

As discussed in Section~\ref{sec:bw_sub}, the Bayesian nature of \bw results in a posterior distribution of the glitch time series.
Previous studies of both simulated signals~\cite{Ghonge2020} and real glitches~\cite{Hourihane:2022doe} have shown that the accuracy of \bw is related to the duration of the data analyzed; \bw more accurately reconstructs short-duration ($\lesssim 1\ {\rm sec}$) events.
To produce the glitch-subtracted data, we use a fair draw from the posterior distribution. An alternative choice is to use the median of the glitch time series posterior, however in some cases this can lead to oversubtraction~\cite{Hourihane:2022doe}.
An example of the 90\% credible interval of the glitch time series posterior for GW190424, along with the whitened glitch model (a fair draw from the posterior), is shown in Figure~\ref{fig:glitch_errors}.  We see that although the fair draw is a good representation of a possible glitch model, we could get slightly different results depending on which glitch time series we happen to draw.

Statistical errors in the glitch model from \gs come from both chance correlations between the two data streams considered and errors in the transfer function measurement itself. 
As described in Section~\ref{sec:lin_sub}, the errors from chance correlations can be calculated directly from parameters used in the glitch modelling. 
In the case of GW200129 the $1\sigma$ statistical errors from chance correlations should be $\frac{1}{\sqrt{2048}} \approx 0.022$. 
The transfer function measurement errors, however, are not directly estimated as part of the glitch modelling process. 
However, if we assume that the transfer function is stationary over a timescale longer than used to measure the transfer function, we can take multiple measurements to estimate this source of error. 

In order to qualitatively estimate the significance of measurement errors, we recalculate the model of the glitches overlapping GW200129 after changing the parameters used in \gs. 
Specifically, we shift the time window used to estimate the transfer function by \result{$\{-1024, -512, 512, 1024\}$ seconds} with respect to original run. 
The glitch model estimated for these four additional attempts is shown in Figure~\ref{fig:glitch_errors} compared against the original model. 
The statistical errors on the original glitch model due to chance correlations is represented by the orange shaded region. 
All five runs show qualitative agreement in the amplitude and phase of the glitch, within the expected errors from chance correlations. 
We therefore conclude that the statistical errors from the glitch modelling procedure are low and should not introduce biases in the analysis. 

It is important to note that these estimates of the errors do account for systematic errors due to modelling assumptions. 
Although the wavelet model used by \bw aims to be fully generic, this model, along with specific choices on the priors used in the analysis, is not able to fully model all glitch features for all glitches. 
Furthermore, specific glitch classes are known to be more problematic for \bw than others~\cite{Hourihane:2022doe}.
Likewise, \gs is limited by the assumption that the witness sensor used to subtract the excess noise is a perfect witness for the noise source and that the transfer function between the witness sensor and the gravitational-wave strain is linear over the amplitudes of interest. 
These types of systematic errors are mitigated by investigations into the limitations of each glitch subtraction algorithm and the choice of which algorithm to apply in each scenario. 
This type of choice was one of the reasons that \gs was used for one of the events in O3. 
Other problematic glitches impacting events in O3 were investigated using \gs, but only the glitches around GW200129 were successfully mitigated with this algorithm.

The potential impact of unaccounted systemic error sources on parameter estimation analyses is not addressed in this work. 
However, studies of \bw-based parameter estimation suggest that systematic errors are low relative to other sources of error in estimation of gravitational-wave event source properties~\cite{Hourihane:2022doe}.
Systematic uncertainties in the case of \gs are more difficult to probe, as these will vary with each noise source considered. 
Previous investigations into the use of \gs and other tools for broadband subtraction did not identify biases~\cite{Davis:2019, Driggers:2018gii}, but it is not clear if these studies are fully applicable in this case due to this differences explained in Section~\ref{sec:lin_sub}. 
As the sensitivity of the detectors increases, these systematic uncertainties may have a larger impact on the measured source properties. 
Recent investigations into GW200129 have already shown that the evidence of orbital precession from this event may be impacted by systematic errors in the glitch modeling~\cite{Payne:2022}.

\section{Future outlook}\label{sec:conc}

The methods described in this work have been used to subtract glitches nearby \result{16} of the gravitational-wave events identified by LVK analyses in O3, representing $\approx 20\%$ the total population to date. 
The high rate of glitches and gravitational-wave signals means that many yet-to-be identified signals will overlap glitches as well. 
The need for this type of mitigation will only grow as the rate of discovery increases in future observing runs~\cite{Aasi:2013wya}.
It is therefore important to consider glitch mitigation and subtraction as an inherent component of gravitational-wave data analysis. 

As the detection rate grows further, there will be a need for additional automation in both the glitch subtraction algorithms and the process for deciding on their use and configuration.
O3 represents the first time that procedures to identify the need for glitch subtraction were utilized due to the high rate of both glitches and signals. 
The need for analysis of more events in O4 will only reduce the amount of person power available on a per-event basis. 
One of the biggest foreseen challenges for future observing runs is reducing human input required to generate glitch-subtracted data.
The procedures and methods described in this work will form a basis for the evaluation of events in future observing runs. 
This effort will also be significantly helped by investigations into the impact of glitches on specific gravitational-wave analyses and additional automation. 

One key area of research that is required to be confident in the astrophysical implications of gravitational-wave signals overlapping glitches is the systematic uncertainties and limitations of different glitch subtraction methods. 
Further complicating these efforts is the potential variability between glitches of the same source, which makes it difficult to apply investigations from past observing runs to new gravitational-wave events. 
Although challenging to estimate, these systematic uncertainties may have significant impact on the conclusions derived from individual events as the sensitivity of the detectors increase.
We therefore urge caution in interpreting results that are dependent on glitch subtraction algorithms when the potential systematics are not fully understand. 
Although this work and previous studies have indicated that such systematics are qualitatively small for tools similar to those used in O3~\cite{Hourihane:2022doe, Driggers:2017, Vajenta:2020ml}, this may not be true for all analyses~\cite{Payne:2022}. 
It is also important to note, however, that while these previous studies have shown that these tools have statistically unbiased results, this does not mean that analyses of individual events will not be impacted. 

The case studies considered in this work also demonstrate the benefit of multiple glitch subtraction tools.
In addition to the personpower limitations created by relying upon a single tool, the wide variety of glitches and glitch-signal overlaps that are possible mean that it is unlikely that a single glitch subtraction method will be best suited for every scenario. 
Even tools that are limited in scope can be useful, as was demonstrated by GW200129.
Additional glitch subtraction methods also could be used as a potential cross-check to understand potential systematic due to specific glitch subtraction tools, as was done in~\cite{Payne:2022}. 
The authors therefore encourage additional investigation into alternative, targeted glitch subtraction methods.
However, new tools are only optimally beneficial when their effectiveness is compared to current methods to understand the specific cases in which one tool is better suited than another. 

As previously described, there are a wide variety of additional glitch subtraction methods that are already available~\cite{Merritt:2021xwh,Driggers:2012noise,Meadors:2014,Tiwari:2015,Driggers:2017,Was:2020ziy,T2100058,VIRGO:2021kfv,Vajenta:2020ml,Ormiston:2020ele,Mukund:2020lby,Yu:2021swq,Mogushi:2021deu,KAGRA:2022frk} or may be ready for use by the next observing run. 
There are also updates already implemented to \bw~\cite{Chatziioannou:2021ezd, Hourihane:2022doe} that were not available for use in O3. 
However, the overall landscape for the data quality of the detectors and the need for glitch subtraction is expected to remain unchanged, as the sources of numerous classes of glitch remain unknown~\cite{LIGO:2021ppb,Acernese:2022jes,KAGRA:2020agh}. 
With the next observing run planned to start in the coming year~\cite{Aasi:2013wya}, now is the time for identifying lessons learned from the O3 glitch subtraction experience and developing new methods to address current limitations.

\section{Acknowledgments}

The authors thank the LIGO-Virgo-KAGRA Detector Characterization and Parameter Estimation groups for their input and suggestions during the development of this work.
We thank Sophie Hourihane and Katerina Chatziioannou for discussions about \bw. 
We also thank Ronaldas Macas for their comments during internal review of this paper.
DD is supported by the NSF as a part of the LIGO Laboratory. IMR-S acknowledges support received from the Herchel Smith Postdoctoral Fellowship Fund.

This material is based upon work supported by NSF’s LIGO Laboratory 
which is a major facility fully funded by the 
National Science Foundation.
LIGO was constructed by the California Institute of Technology 
and Massachusetts Institute of Technology with funding from 
the National Science Foundation, 
and operates under cooperative agreement PHY-1764464. 
Advanced LIGO was built under award PHY-0823459.
The authors are grateful for computational resources provided by the 
LIGO Laboratory and supported by 
National Science Foundation Grants PHY-0757058 and PHY-0823459.
This work carries LIGO document number P2200192.

Parts of this research are supported by the Australian Research Council (ARC) Centre of Excellence for Gravitational
Wave Discovery (OzGrav) (project number CE170100004)
and ARC Discovery Project DP170103625.

\appendix

\section{Linear subtraction tests with simulated data}\label{sec:sim_data}

\begin{figure}[t]
  \centering
  \includegraphics[width=\textwidth]{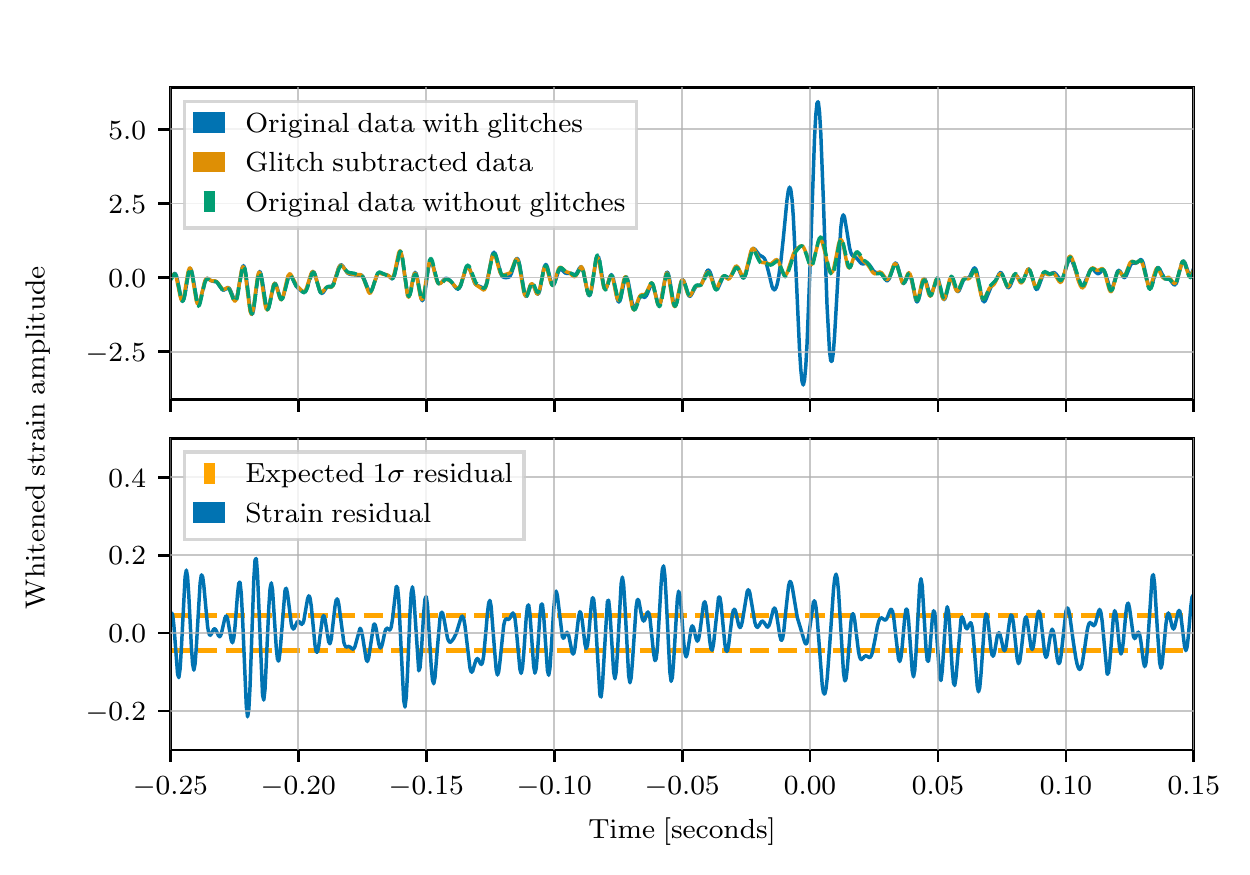}
  \caption{A comparison of whitened time series for simulated data (Top) with a glitch in blue, simulated data with a glitch subtracted in orange, and simulated data that does not contain the glitch in green.
  The similarity of the three time series for times far away from the glitch show that there was correctly only minimal subtraction done for these time periods. 
  The residual between the simulated data with a glitch subtracted and the simulated data that does not contain the glitch (Bottom) is also shown in blue. The expected standard deviation of the residual is $\approx 0.05$, which roughly agrees with the measured standard deviation of $0.068$. 
  }
  \label{fig:sim_data}
\end{figure}

As \gs has not been previously used to intentionally subtract glitches from gravitational-wave strain data,
it is important to confirm that the algorithm can be used in this scenario. 
While glitches should theoretically be subtracted using the procedures described in Section~\ref{sec:lin_sub}, the practical differences listed in this section may make glitch subtraction difficult. 
Cases where glitches were subtracted were identified in investigations of the data produced in~\cite{Davis:2019}, but the glitch subtraction effectiveness was not assessed. 
To confirm that \gs could be used to subtract glitches in a controlled setting, we test the algorithm with simulated data.

For this test, we first generate a time series with sine-Gaussian wavelets and a low amplitude background (i.e. an auxiliary witness) and a time series that includes both Gaussian noise colored to match the spectrum of LIGO Livingston in O3 and the same-Gaussian wavelets (i.e. the gravitational-wave strain).
The sine-Gaussian wavelets in each channel are related by a known transfer function. 
We then process the simulated strain data with \gs to generate a glitch-subtracted time series.
A comparison of the original strain data with glitches and the glitch-subtracted data is shown in the top panel Figure~\ref{fig:sim_data}.
Alongside these time series, we also plot the original data without the sine-Gaussian glitches included. 

Ideally, the glitch-subtracted data should match the original data without glitches, up to expected statistical uncertainties.
The difference between these two time series is shown in the bottom panel of Figure~\ref{fig:sim_data}.
With the settings used in this analysis, the residual should have standard deviation of $\frac{1}{512} \approx 0.05$.
The true standard deviation of the residual is 0.068, which qualitatively agrees with expectations.  

\section{PE studies of glitch subtraction}\label{sec:pe}

\begin{figure}[p]
    \centering
    \includegraphics[width=0.48\textwidth]{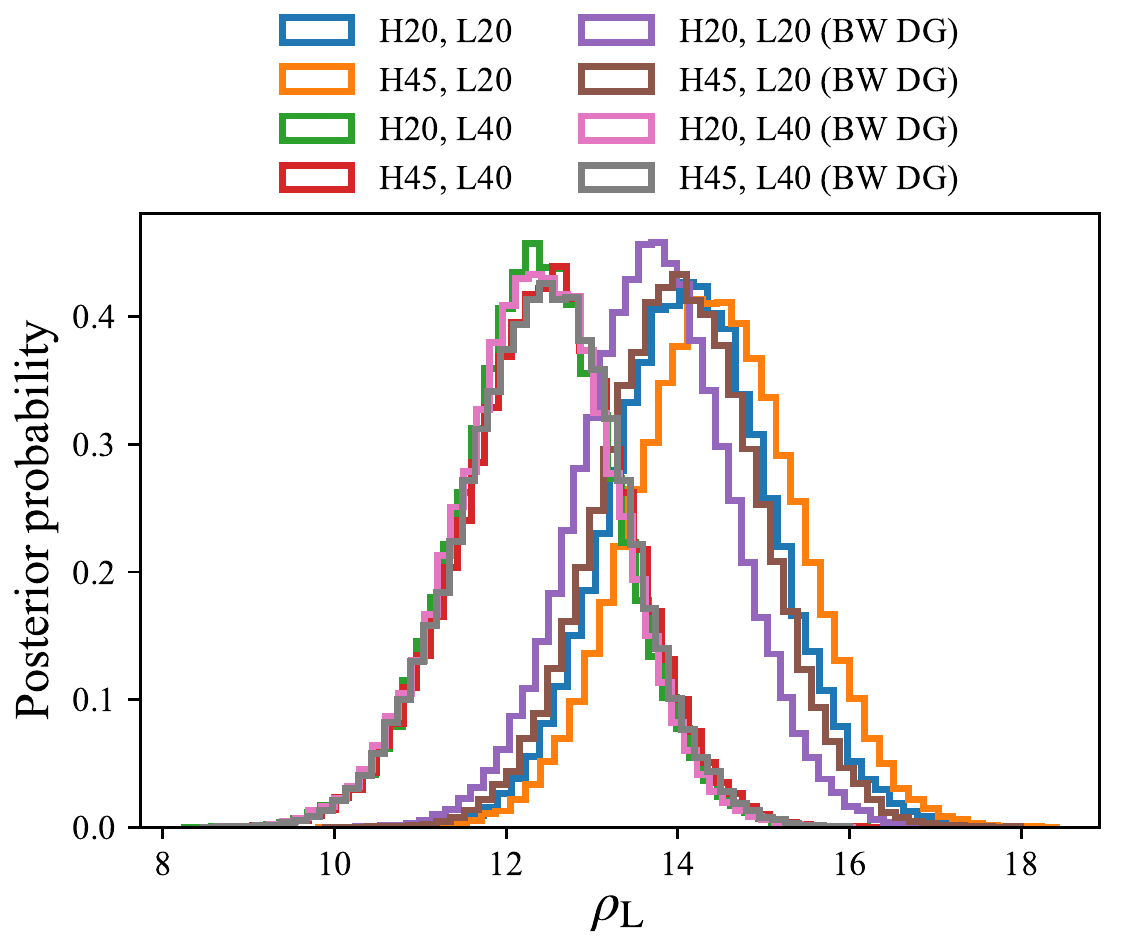}
    ~
    \includegraphics[width=0.48\textwidth]{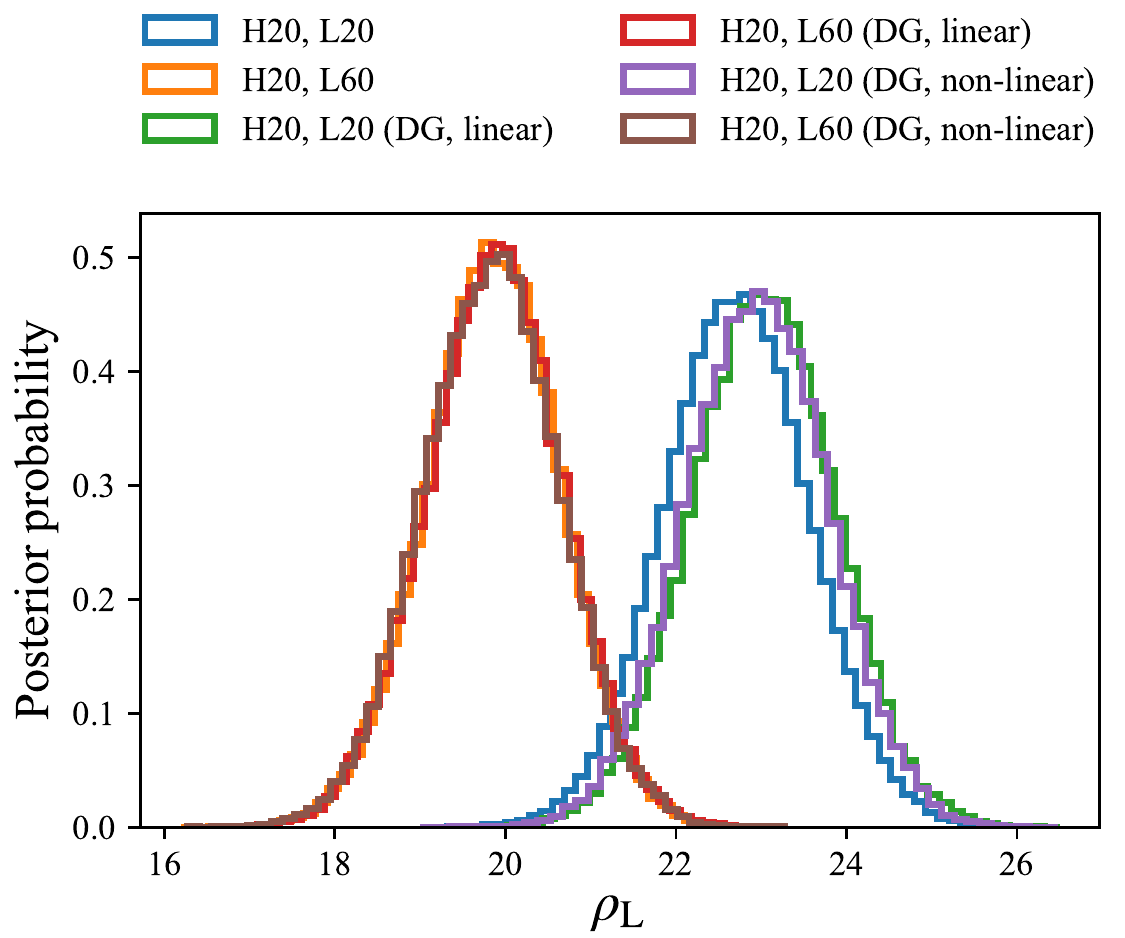}
  \includegraphics[width=0.48\textwidth]{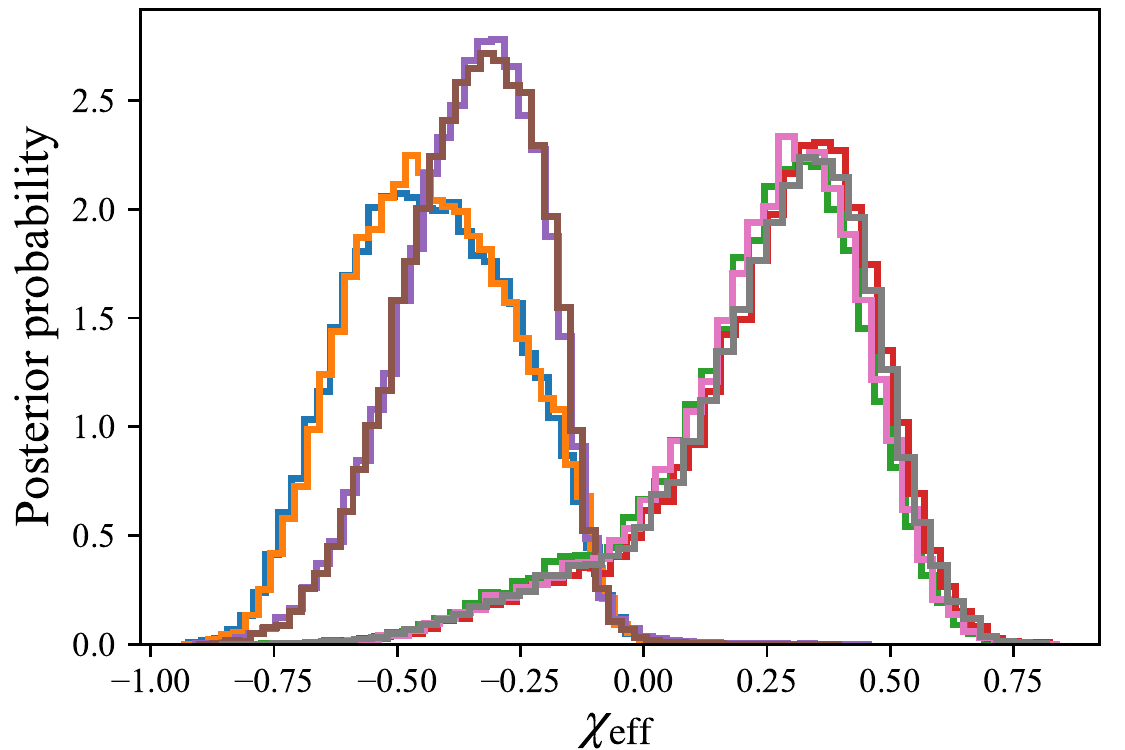}
      ~
    \includegraphics[width=0.48\textwidth]{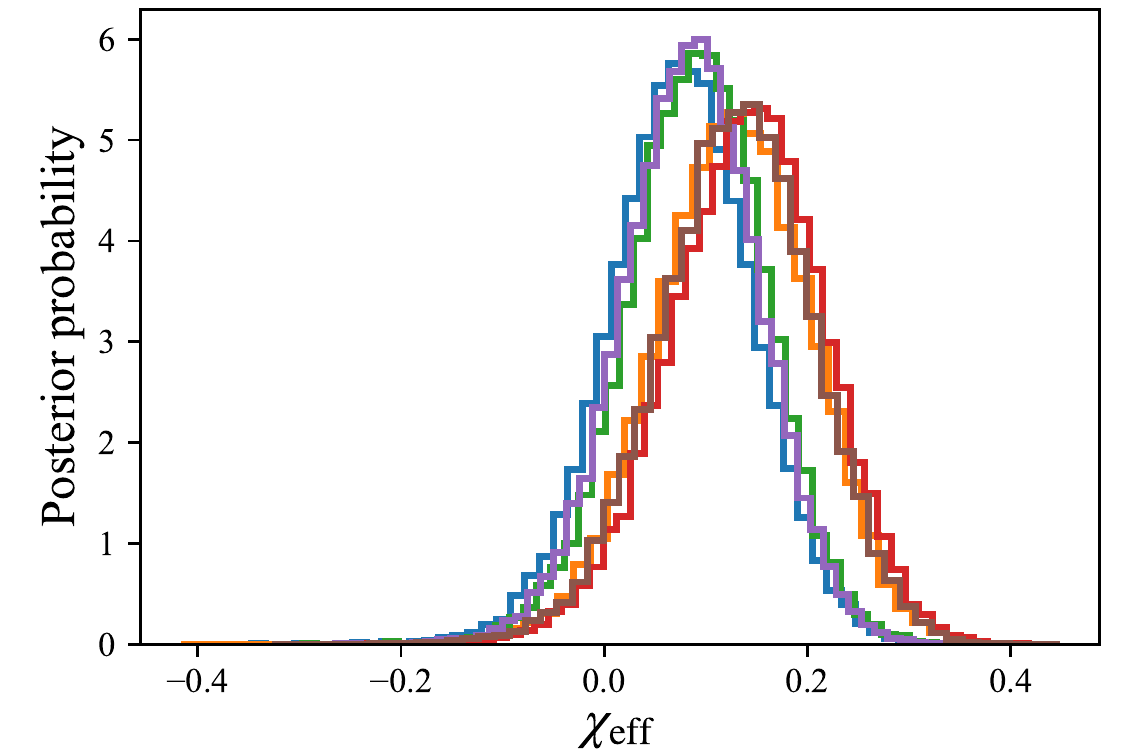}
  \includegraphics[width=0.48\textwidth]{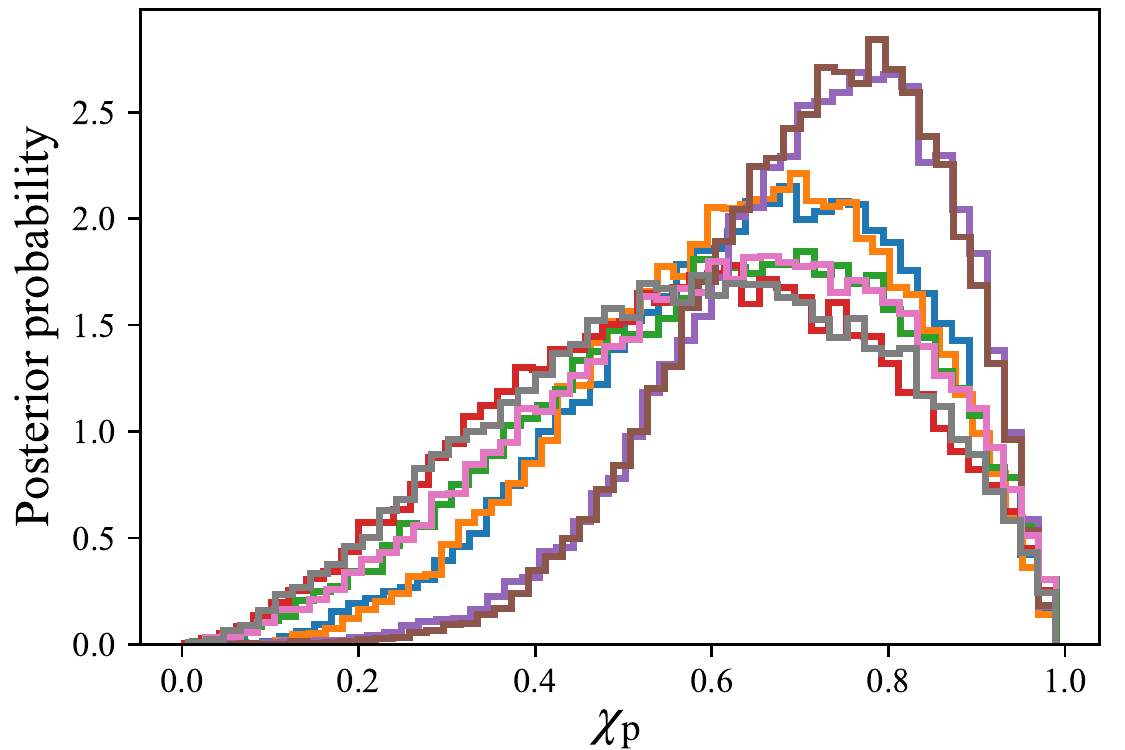}
      ~
    \includegraphics[width=0.48\textwidth]{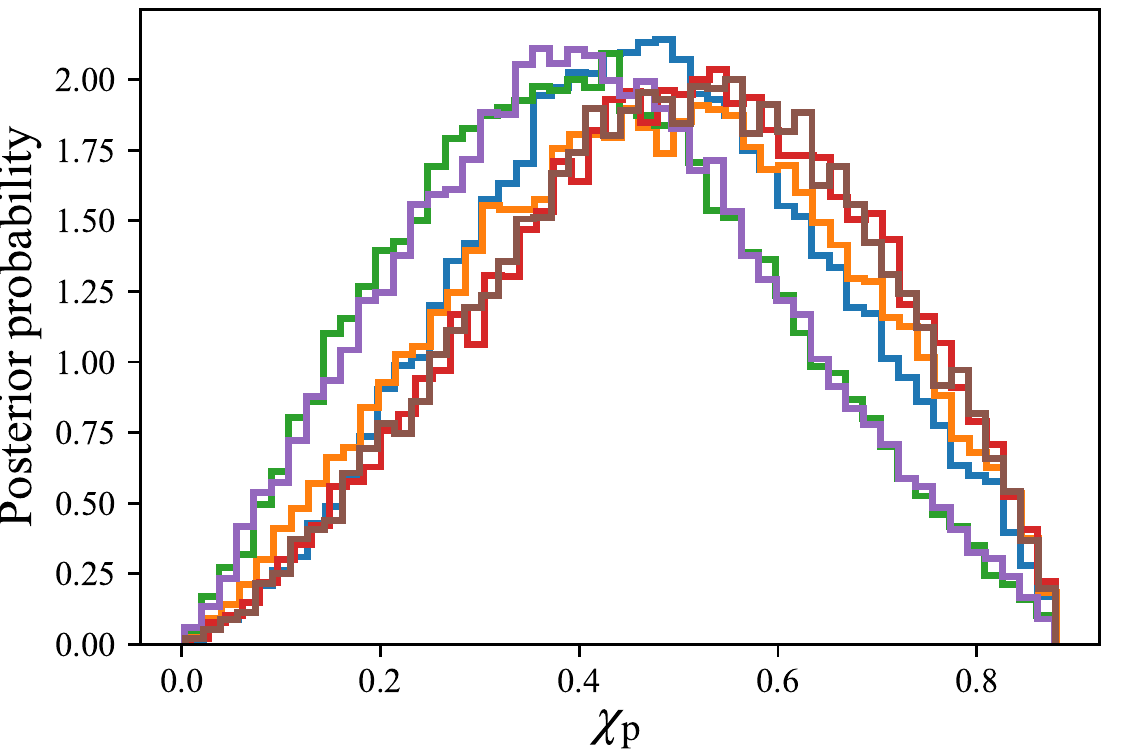}
  \caption{Plots showing the posteriors on matched-filter SNR in the LIGO Livingston detector $\rho_L$ (Top), effective spin parameter $\chi_\mathrm{eff}$ (Middle), and effective precession parameter $\chi_\mathrm{p}$ (Bottom) under different mitigation strategies for GW191109 (Left) and GW200129 (Right). The unmitigated posterior is shown in blue in all plots. The ``successful'' posterior (representing the chosen mitigation strategy) is shown in purple for GW191109 (Left; ``H20, L20 (BW dg)'') and green for GW200129 (Right; ``H20, L20 (dg, linear)''). The Livingston SNR is highest when the nominal lower frequency of $20$\,Hz is used for the Livingston data. For GW191109, \bw deglitching preserves the support for negative $\chi_\mathrm{eff}$, while raising the lower frequency at Livingston to $40$\,Hz pushes the majority of the support to positive values. In both cases, the most informative posteriors come from deglitched data with the nominal lower frequency limit of $20$\,Hz. }
  \label{fig:pe_compare}
\end{figure}

For events with relatively low-frequency glitches, there are two potential data mitigation strategies: (a) using only data above a certain cut-off frequency, below which the glitch resides; or (b) modelling and subtracting the glitch. To converge on the best approach to use for an event, we perform Bayesian inference on real data with various mitigation strategies applied. For each event, we then visually compare the posterior probability distributions for the network and per-detector signal-to-noise ratios (SNRs) and other parameters of interest to inform our choice of mitigation strategy. In every case for which these PE studies were performed, glitch subtraction was chosen as the preferred mitigation strategy. 

We use \texttt{Bilby} and \texttt{bilby\_pipe}~\cite{Ashton:2018jfp, Romero-Shaw:bilby:2020} to perform our parameter estimation and obtain posterior probability distributions on binary parameters. Aside from changing the data or lower frequency limits, the remainder of the settings are identical between runs on the same event. We use the \texttt{IMRPhenomPv2} waveform approximant~\cite{Husa:2015iqa,Khan:2015jqa}, a sampling frequency of $2048$\,Hz, a maximum frequency of $1024$\,Hz, and a reference frequency of $20$\,Hz. Unless altered as part of a mitigation strategy, the default lower frequency of analysis is $20$\,Hz. We employ marginalisation in phase, time and distance. 

Here we focus on the results of two case studies: GW191109\_010717 (Henceforce referred to as GW191109) and GW200129.\footnote{The other events that were subject to PE comparison studies were GW190413\_134308, GW190424\_180648, GW190513\_205428, GW190514\_065416, GW190727\_060333, GW190828\_065509, GW190910\_112807, GW190915\_235702, GW191105\_143521, GW191127\_050227, and GW200115\_042309.} For both of these events, we perform analyses on $4$\,s of data.
Data from LIGO Hanford and LIGO Livingston was used for both events, while Virgo data was included for runs with GW200129 (Virgo was offline at the time of GW191109).
We use \texttt{bilby} default priors in both cases: the \texttt{high\_mass} prior for GW191109, and the $4$\,s prior for GW200129 (see \cite{Romero-Shaw:bilby:2020} for details of these priors).

GW191109 is one of the most likely events in O3 to exhibit signs of negatively-aligned spins. 
For GW191109, the data sets that we test each employ of the following mitigation strategies:
\begin{enumerate}
    \item \textbf{``H45, L20"}: Limiting the frequency of Hanford data to above $45$\,Hz;
    \item \textbf{``H20, L45"}: Limiting the frequency of Livingston data to above $40$\,Hz;
    \item \textbf{``H45, L40"}: Limiting the frequency of Hanford and Livingston data to above $45$\,Hz and $40$\,Hz respectively;
    \item \textbf{``BW dg"}: \bw glitch subtraction.
\end{enumerate}

GW200129 is one of the most likely events in O3 to contain evidence for spin-induced orbital precession. 
For GW200129, the compared data sets each individually have one of the following mitigation strategies applied:
\begin{enumerate}
    \item \textbf{``H20, L60"}: Limiting the frequency of Livingston data to above $60$\,Hz;
    \item \textbf{``dg, non-linear"}: Linear glitch subtraction on data previously cleaned with broadband non-linear $60$\,Hz subtraction;
    \item \textbf{``dg, linear"}: Linear glitch subtraction on data previously cleaned with broadband linear $60$\,Hz subtraction.
\end{enumerate}
We also compare these results against results obtained using unmitigated data (``H20, L20''). 

Results are shown for matched-filter Livingston SNR, effective aligned spin $\chi_\mathrm{eff}$, and effective precession spin $\chi_\mathrm{p}$ in Figure \ref{fig:pe_compare} for GW191109 and GW200129. 
We find that when we increase the lower-frequency limit of analysis for LIGO Livingston data, the Livingston optimal SNR is reduced, as expected.
Meanwhile, there is little change to the SNR when the various glitch subtraction strategies are employed.
There is also negligible difference between posteriors obtained with the linearly and non-linearly cleaned data for GW200129.

The posteriors on spin parameters $\chi_{\rm eff}$ and $\chi_{\rm p}$ are dependent on our choice of data mitigation strategy.
When the lower frequency limit is raised above $20$\,Hz, information about the spins of the system can be lost, as less of the inspiral track is contained within the reduced extent of the data.
Therefore, posteriors on spin parameters with limited low-frequency data are more influenced by the prior and are less well-constrained.
In both cases, deglitching the data narrows the posteriors on $\chi_{\rm eff}$ and $\chi_{\rm p}$ relative to the posteriors computed from unmitigated data and reduced low-frequency data.

Low frequencies contain most of the information about $\chi_\mathrm{p}$, so the red histogram in both plots on the final row roughly traces the prior (uninformative) distribution on $\chi_\mathrm{p}$; both events have more informative $\chi_\mathrm{p}$ posteriors when deglitched than in the unmitigated case.
In the case of GW191109, it appears that Hanford contains very little information about spin parameters: the $\chi_{\rm eff}$ and $\chi_{\rm p}$ posteriors with the Hanford lower frequency limit at $45$\,Hz and no deglitching (orange) follow the posteriors with no data mitigation strategy applied (blue).
Since the SNR loss was significant when higher limits were placed on the lower frequency, our choice for both events demonstrated here was to deglitch the data and maintain a lower frequency limit of $20$\,Hz.

\section*{References}
\bibliography{main.bbl}

\end{document}